\documentclass[twocolumn, aps, pra, nofootinbib, nobibnotes, showpacs]{revtex4}

\usepackage{amsmath}   
\usepackage{amsthm}    
\usepackage{amssymb}   

\newcommand \conj	{{}^\ast}        
\renewcommand \ge	{\geqslant}      
\renewcommand \le	{\leqslant}      
\newcommand \x		{\times}         
\newcommand \C		{\mathbb C}      
\newcommand \e		{\mathrm e}      
\newcommand \tr		{\text{\textrm{\textup{tr}}}}  
\renewcommand \implies	{\Longrightarrow}             
\renewcommand \iff		{\Longleftrightarrow}         
\newcommand \ox		{\otimes}                     
\newcommand \inv	{^{-1}}          
\newcommand \sgn        {\operatorname{sgn}}            

\newcommand \smallto	{\rightarrow}					  
\renewcommand \mapsto	{\longmapsto}					  
\renewcommand \epsilon	{\varepsilon}					  

\newcommand  \boost[1]	{\big. #1 \big.}                       
\renewcommand\brace[1]	{\left\{ #1 \right\}}		       
\newcommand  \abs[1]	{\left| #1 \right|}		       
\newcommand  \ket[1]	{\boost{\left| #1 \right\rangle}\!}    
\newcommand  \bra[1]	{\boost{\left\langle \!\!\: #1 \right|}}      
\newcommand  \bracket[2]{\boost{\left\langle \!\!\: #1 \!\left.\right|\!\!\: #2 \right\rangle}}

\newcommand  \ketbra[2]	{\boost{\left| #1 \!\!\;\left\rangle\!\right\langle\!\!\: #2 \right|}}
\newcommand  \densmtx[1]{\ketbra{#1}{#1}}		       
\newcommand  \paren[1]	{\left( #1 \right)}		       
\newcommand  \sqparen[1]{\left[ #1 \right]}		       

\newcommand \sss[1]	{ \scriptscriptstyle{#1} }

\renewcommand\qed{\hfill$\blacksquare$}

\newtheoremstyle{theorem}{6pt}{3pt}{\itshape}{}{\bfseries}{.}{1em}
		{\thmname{#1}\thmnumber{ #2}\thmnote{ (#3)}}
		
\theoremstyle{theorem}
\newtheorem{theorem}{Theorem}

\newtheorem{lemma}{Lemma}

\newtheorem{corollary}{Corollary}[theorem]

\newtheoremstyle{theoremname}{6pt}{3pt}{\itshape}{}{\bfseries}{.}{1em}
		{\thmname{#3}}

\theoremstyle{theoremname}

\newtheoremstyle{dfn}{6pt}{3pt}{}{}{\bfseries}{.}{1em}{}
\theoremstyle{dfn}

\newtheorem{definition}{Definition}

\newtheoremstyle{example}{6pt}{3pt}{}{}{\itshape\bfseries}{.}{1em}
		{\thmname{#1}\thmnumber{ #2}\thmnote{: \textmdit{#3}}}
\theoremstyle{example}

\newtheoremstyle{examplecontd}{6pt}{3pt}{}{}{\bfseries\itshape}{}{1em}
		{\thmname{#1}\thmnumber{ #2}\thmnote{ #3\textmdit{\textbf{.} (cont'd)}}}

\newtheoremstyle{axiom}{6pt}{3pt}{\itshape}{ }{\scshape}{.}{1em}{}
\theoremstyle{axiom}

\newtheoremstyle{axiomname}{6pt}{3pt}{\itshape}{ }{\scshape}{.}{1em}
		{\thmname{#3}}
\theoremstyle{axiomname}

\newtheoremstyle{notation}{6pt}{3pt}{}{}{\bfseries}{.}{1em}{\thmname{#1}}
\theoremstyle{notation}

\newtheoremstyle{convention}{6pt}{3pt}{}{}{\itshape}{:}{1em}{\thmname{#1}}
\theoremstyle{convention}

\newtheoremstyle{proof}{0mm}{3pt}{}{}{\bfseries\itshape}{ ---}{1em}
		{\thmname{#1}\thmnumber{ #2}\thmnote{ #3}}
\theoremstyle{proof}
\newtheorem*{pf}{Proof}
\renewenvironment{proof}[1][]
		 {\begin{pf}[#1]}
		 {\qed\end{pf}}

\renewcommand \a   {{\sss \alpha}}
\renewcommand \b   {{\sss \beta}}
\newcommand \s   {{\sss \sigma}}
\newcommand \mP  {{\mathcal P}}
\newcommand \mQ  {{\mathcal Q}}
\newcommand \cswap {{\,\textsc{c-swap}\,}}

\begin{document}

\title{One-qubit fingerprinting schemes}
\author{J. Niel de Beaudrap}
\thanks{Email: {\tt jd@cpsc.ucalgary.ca}}
\affiliation{Department of Computer Science,
		 University of Calgary, Calgary, Alberta, Canada T2N 1N4}

\date{\today}

\begin{abstract}
Fingerprinting is a technique in communication complexity in which two parties
(Alice and Bob) with large data sets send short messages to a third party (a
\emph{referee}), who attempts to compute some function of the larger data sets.
For the equality function, the referee attempts to determine whether Alice's data
and Bob's data are the same. In this paper, we consider the extreme scenario of
performing fingerprinting where Alice and Bob both send either one bit (classically)
or one qubit (in the quantum regime) messages to the referee for the equality problem.
Restrictive bounds are demonstrated for the error probability of one-\emph{bit}
fingerprinting schemes, and show that it is easy to construct one-\emph{qubit}
fingerprinting schemes which can outperform any one-bit fingerprinting scheme. The
author hopes that this analysis will provide results useful for performing physical
experiments, which may help to advance implementations for more general quantum
communication protocols.
\end{abstract}


\maketitle


\section{Introduction}

Fingerprinting is a useful mechanism for computing functions of large and distributed
sets of data, by sending short messages called `fingerprints' of the data to a third
party for computation. The aim is to reduce the amount of communication, at the cost
of some manageable probability of error.

The notion of fingerprinting arises naturally in the setting of \emph{communication
complexity} (see \cite{KN97}). In the \emph{simultaneous message passing model}, which
was introduced by Yao in his original paper on communication complexity \cite{Yao79},
two parties (Alice and Bob) each have a piece of data represented by a string
($\alpha$ for Alice and $\beta$ for Bob). They wish to compute some function $f$ of
their strings, $f(\alpha,\beta)$. However, we constrain Alice and Bob in the following
way: they may not communicate with one another, but they may each send one message to
a third party (a \emph{referee}) who will attempt to compute $f(\alpha,\beta)$ using
messages sent by Alice and Bob. It is assumed that Alice, Bob, and the referee have
established some protocol which describes:
\begin{itemize}
\item how Alice and Bob choose messages to send to the referee depending on their
      strings, and
\item how the referee interprets the messages in order to attempt to compute
      $f(\alpha,\beta)$.
\end{itemize}
In our scenario, we are assuming Alice and Bob do not have access to a common source of
random bits. A simple protocol in which Alice and Bob simply transmit (a binary
representation of) their entire strings will suffice. The question that is asked is: how
great a probability of success can they attain if they each transmit less information to
the referee? We refer to the shorter messages that Alice and Bob send as \emph{fingerprints}
of their data. In the case where Alice and Bob wish to compare their strings for equality,
we may use $f(\alpha,\beta) = \delta_{\a\b}\,$, which evaluates to $0$ if $\alpha = \beta$
and $1$ otherwise.

Performing fingerprinting using quantum information was first investigated for
the asymptotic case in \cite{BCWW01}. In that paper, it was shown that fingerprints
of length $O(\log n)$ suffice for determining whether or not two $n$-bit strings
are equal, with an error probability less than $\epsilon$ (for any constant $\epsilon > 0$).
This contrasts with the classical case, where it is known that fingerprints 
of size $\Omega(\sqrt n)$ are necessary for comparable performance \cite{NS96}.
The result of \cite{BCWW01} is part of a sequence of results including \cite{BCW98, Raz99},
where it has been shown that communication complexity can be reduced using quantum
communication. Although these results demonstrate asymptotic savings, they do not
explicitly demonstrate savings in small instances that might be suitable for experimental
implementations in order to verify the theory. For example, it is not clear that
the significant quantum/classical communication separations shown in \cite{BCW98}
and \cite{Raz99} hold for particularly small problem-size instances.

In this paper, we examine very small instances of fingerprinting. Specifically,
if Alice and Bob may only transmit one bit (or one qubit) of information to the referee,
what can we say about the probability of success in determining whether $\alpha = \beta$\,?
By analysing these very limited instances of the quantum fingerprinting problem, we hope to
provide results which will be easily tested by experiment, in order to help the
development of implementations for more general quantum communication protocols.

It should be noted that Massar \cite{Mas03} proposed a single \emph{particle}
fingerprinting scheme, in which a particle is sent either to Alice or to Bob
(in superposition). Each party induces a relative phase on the state of the particle,
after which the two paths are interfered with one another. This scenario differs
from the one considered here --- for example, Alice and Bob share quantum information
(the path taken by the particle) in addition to their input data ($\alpha$ and $\beta$),
whereas no such additional information is available in the present scenario.

\section{Preliminary Definitions}

Let $S$ be the set of possible strings that Alice and Bob may have. We consider cases
where $\abs{S} > 2$ (cases with $\abs{S} \le 2$ are of course trivial to solve with
one bit of communication).

\begin{definition}
Given the one (qu)bit messages from Alice and Bob, the referee will make
a declaration (correct or incorrect) about whether or not their strings match.
\begin{itemize}
\item
If the referee declares that the strings match, we say that his decision strategy
\emph{accepts} the input it was given, and we call this a \emph{positive} result;
if this declaration is incorrect, we may call this a \emph{false positive}. 

\item
If the referee declares that the strings differ, we say that his decision strategy
\emph{rejects} the input it was given, and we call this a \emph{negative} result;
if this declaration is incorrect, we may call this a \emph{false negative}.
\end{itemize}
\end{definition}

\begin{definition}
Here are some definitions related to error probabilities.
\begin{itemize}
\item
For a given fingerprinting scheme, the \textit{error probability} of the scheme will
refer to the worst-case probability of an incorrect result in that scheme.

\item
A fingerprinting scheme with \textit{one-sided error} is one where, if Alice and Bob
have the same data string, the probability that the scheme will produce a false
negative is zero.
\end{itemize}
\end{definition}

\section{Classical one-bit fingerprinting}
\label{sxn:classical}

In this section, we provide the definition of a one-bit fingerprinting scheme,
and show that such schemes have quite limited power.
Specifically, any one-bit fingerprinting scheme has an error probability of
at least $\frac{1}{2}$ (which is the same error probability that results from 
just a fair coin flip by the referee).
As well, any one-bit fingerprinting scheme with one-sided error has error 
probability 1 (the same error probability that results if the referee always 
declares that the strings match).
Finally, any one-bit fingerprinting scheme that succeeds with high probability
when Alice's and Bob's strings match necessarily fails with high probability in
$\Theta(\abs{S}^2)$ of the cases where their strings do not match.

\begin{definition}
A \textit{one-bit fingerprinting scheme} $\mP$ operates as follows.
To begin with, Alice and Bob receive inputs $\alpha, \beta \in S$, respectively.
Each of them sends one bit to the referee, who produces an output bit based on
these two messages. Each party may use a probabilistic strategy; however, there is no 
\textit{shared} randomness among the parties.
Such a scheme is characterised by three probability distributions:
\begin{itemize}
\item
Alice's fingerprinting strategy $\brace{p_\s}_{\sigma \in S}$\,, where $p_\a$ is the
probability with which Alice sends the message {\bf 1}, depending on her string
$\alpha \in S$\,.
\item
Bob's fingerprinting strategy $\brace{q_\s}_{\sigma \in S}$\,, where $q_\b$ is the probability
with which Bob sends the message {\bf 1}, depending on his string $\beta \in S$\,.
\item
The referee's decision strategy $\brace{r_{00}, r_{01}, r_{10}, r_{11}}$, 
where $r_{ab}$ is the probability with which the referee declares a positive 
result, depending on the bits $a$ and $b$ sent to him by Alice and Bob 
respectively.
\end{itemize}
\end{definition}

\begin{definition}
For any one-bit fingerprinting scheme $\mP$, define the function
\begin{align}
\label{defn:P_plus}
   \begin{split}
        \mP_+(\alpha, \beta) \;=\;&
		(1-p_\a)(1-q_\b)r_{00}	\;+\;
		(1-p_\a) q_\b r_{01}	\\&\;+\;
		p_\a (1-q_\b) r_{10}	\;+\;
		p_\a q_\b r_{11} \,.
   \end{split}
\end{align}
That is, $\mP_+(\alpha, \beta)$ is the probability that the scheme $\mP$ yields
a positive result when the input to Alice is $\alpha$ and to Bob is $\beta$\,.
\end{definition}

We will usually want to place lower bounds on the probability of a positive
result when $\alpha = \beta\,$. This will have a strong impact on the overall
performance of the scheme $\mP$. In order to simplify analysis, it will be useful
to come up with a simple inequality.

\begin{lemma}
\label{thm:clasIneq}
For any one-bit fingerprinting scheme $\mP$, it is possible to define real-valued
parameters $x_\a\,$, $y_\b\,$, and $d_\epsilon$ (depending on $\alpha, \beta \in S$ and
$\epsilon \in \sqparen{0,1}$ respectively) such that
\begin{align}
\label{eqn:clBound}
	x_\a \: y_\b \;\ge\; d_\epsilon
	    \quad&\iff\quad
	\mP_+(\alpha, \beta) \;\ge\; (1 - \epsilon)\,.
\end{align}
\end{lemma}
\begin{proof}
Let $\alpha, \beta \in S$ be strings such that $\mP_+(\alpha, \beta) \ge (1-\epsilon)$ holds.
We may expand the expression of $\mP_+(\alpha, \beta)$ in Equation~\ref{defn:P_plus}
as a polynomial in $p_\a$ and $q_\b$: this gives us
\begin{align}
   \begin{split}
        \label{eqn:P_plus_poly}
	(r_{00} - &r_{01} - r_{10} + r_{11})  p_\a q_\b	\;+\;
	(r_{10} - r_{00}) p_\a				\\&\;+\;
	(r_{01} - r_{00}) q_\b				\;+\;
	r_{00}			\quad\ge\quad (1-\epsilon) \,.
   \end{split}
\end{align}
We would like to separate the dependency on $\alpha$ and $\beta$: how we may do this
depends on the value of $c = (r_{00} - r_{01} - r_{10} + r_{11})$\,.
\begin{itemize}
\item
If $c \not= 0$, we may define the parameters $x_\a\,$, $y_\b\,$, and $d_\epsilon$
as follows:
\begin{align}
   \begin{split}
	x_\a \:&=\: c\paren{p_\a \,+\, \frac{r_{01} \,-\, r_{00}}{c}} \\
	y_\b \:&=\: q_\b \,+\, \frac{r_{10} \,-\, r_{00}}{c}          \\
	d_\epsilon \:&=\: (1-\epsilon) \,+\, 
		  \frac{(r_{00} \,-\, r_{01})(r_{00} \,-\, r_{10})}{c} \,-\, r_{00}\,.
   \end{split}
\end{align}
Then, the inequality $x_\a\,y_\b \ge d_\epsilon$ holds, as it is equivalent to
Equation~\ref{eqn:P_plus_poly}.

\item
If $c = 0$, we may drop the cross-term in $p_\a$ and $q_\b$ from
Equation~\ref{eqn:P_plus_poly}:
\begin{align}
  \begin{split}
	(r_{10} - r_{00}) p_\a	\;&+\; (r_{01} - r_{00}) q_\b \\ &+\; r_{00}
	\quad\ge\quad (1-\epsilon) \,.
  \end{split}
\end{align}
As the exponential function $(x \,\mapsto\, \e^x)$ is monotone increasing, the inequality
will be preserved if we exponentiate both sides. Doing this yields
\begin{align}
        \e^{(r_{10} - r_{00}) \, p_\a} \; \e^{(r_{01} - r_{00}) \, q_\b} \; \e^{r_{00}}
	\;\ge\; \e^{1 \,-\, \epsilon} \,.
\end{align}
In this case, we may define the parameters $x_\a\,$, $y_\b\,$, and $d_\epsilon$ as follows:
\begin{align}
  \begin{split}
        x_\a \:&=\: \e^{(r_{10} - r_{00}) \, p_\a}  \\
	y_\b \:&=\: \e^{(r_{01} - r_{00}) \, q_\b}  \\
	d_\epsilon \:&=\: \e^{1 \,-\, \epsilon \,-\, r_{00}} \,.
  \end{split}
\end{align}
Then, we have the inequality $x_\a \: y_\b \;\ge\; d_\epsilon$\,.
\end{itemize}
In both cases, the parameters are defined so that if \,$\mP(\alpha,\beta) \,\ge\,
(1-\epsilon)$\,, the inequality $x_\a\:y_\b\;\ge\;d_\epsilon$ holds. For the converse,
suppose that $x_\a\:y_\b\;\ge\;d_\epsilon$ for two particular strings $\alpha, \beta \in S$.
By simply applying the definitions of $x_\a\,$, $y_\b\,$, and $d_\epsilon$ (and taking the
logarithms of both sides if $c = 0$), we can immediately recover
Equation~\ref{eqn:P_plus_poly}. Thus, the converse holds as well.
\end{proof}

The parameters $x_\a$, $y_\b$, and $d_\epsilon$ do not necessarily refer to any
obvious properties of the scheme $\mP$. However, using them, we may easily
prove very restrictive upper bounds on the effectiveness of one-bit fingerprinting
schemes:

\begin{theorem}
\label{thm:classicalLowerBound}
Let $\mP$ be a one-bit fingerprinting scheme. If $\epsilon$ is an upper bound
for the worst-case probability that $\mP$ produces false negatives, then $(1 - \epsilon)$
is a lower bound on the worst-case probability that $\mP$ will produce false positives.
\end{theorem}
\begin{proof}
Because the probability of a false negative is at most $\epsilon$, we
have $\mP_+(\sigma, \sigma) \ge (1-\epsilon)$ for all strings $\sigma \in S$.
Then, by Lemma~\ref{thm:clasIneq}, we also have $x_\s \, y_\s \,\ge\, d_\epsilon$ for
all $\sigma \in S$. Consider an alternative definition of the `sign' function:
\begin{align}
	\sgn x_\s &
	=	\begin{cases}
			\;~1\,, & x_\s \ge 0	\\
			-1\,, & x_\s < 0
		\end{cases} \;\;.
\end{align}
As there are two possible values for the `sign' function, and more than two elements
in $S$, there must then be at least two distinct strings $\mu, \nu \in S$ such that
$\sgn x_\mu = \sgn x_\nu$\ . Without loss of generality, we may choose $\mu$ and $\nu$
such that $(\sgn x_\nu)\:y_\nu \;\ge\; (\sgn x_\mu)\:y_\mu$. Then we have
\begin{align}
	x_\mu\:y_\nu  \;\;&\ge\;\;  x_\mu\:y_\mu \;\;\ge\;\; d_\epsilon \,.
\end{align}
Again by Lemma~\ref{thm:clasIneq}, we then know that $\mP_+ (\mu, \nu) \ge
\hbox{$(1-\epsilon)$}$. As $\mu$ and $\nu$ were chosen to be distinct, the worst-case
probability of a false positive for the fingerprinting scheme $\mP$ is then at least
$(1 - \epsilon)$.
\end{proof}


\begin{corollary}
\label{thm:classicalTwoSide}
Any one-bit fingerprinting scheme has error probability at least $\tfrac{1}{2}$.
\end{corollary}

\begin{corollary}
\label{thm:classicalOneSide}
Any one-sided error one-bit fingerprinting scheme has error probability $1$.
\end{corollary}

Not only are there lower bounds on the worst-case probability of failure, but
there is also a lower bound on the number of inputs which are likely to produce
a false positive:

\begin{theorem}
Let $\mP$ be a one-bit fingerprinting scheme, and let $\abs{S} = s$. If 
the worst-case probability of obtaining a false negative is at most $\epsilon \ge 0$
in $\mP$, then the number of inputs $(\alpha, \beta) \in S \x S$ such that
$\alpha \not= \beta$ and $\mP_+(\alpha, \beta) \;\ge\; (1 - \epsilon)$ is at
least $\tfrac{1}{4}(s^2 - 2s)$.
\end{theorem}
\begin{proof}[sketch]
This follows by considering the arguments from Theorem~\ref{thm:classicalLowerBound}.
We may establish a lower bound by counting the number of unordered pairs
$\{\alpha, \beta\}$ such that $\sgn x_\a = \sgn x_\b$ and $\alpha \ne \beta$.
This is minimized if nearly half of the strings $\sigma \in S$ have $\sgn x_\s = 1$,
and nearly half again have $\sgn x_\s = -1$. If this is the case, a simple counting
argument shows that the number of such (unordered) pairs is
$2 \binom{s/2}{2} = \tfrac{1}{4}(s^2 - 2s)$ for $s$ even, and $\binom{(s+1)/2}{2} +
\binom{(s-1)/2}{2} = \tfrac{1}{4}(s^2 - 2s + 1)$ for $s$ odd.
\end{proof}

So, one is forced to accept one of two scenarios in a one-bit fingerprinting scheme:
one must either forego a reasonable upper bound on the probability of error when Alice's
and Bob's strings match, or use a scheme which will produce false positives with high
probability for about $\tfrac{1}{4}$ or more of Alice's and Bob's possible inputs
(strictly more than the number of matching cases when $\abs{S} > 6$\,).

\section{Quantum one-qubit fingerprinting}

In this section, we provide the definition of a one-qubit fingerprinting scheme,
give examples of one-qubit schemes that outperform what classical one-bit fingerprinting
schemes can achieve, and give some upper and lower bounds for the performance of schemes with
one-sided error.

\begin{definition}
\label{dfn:quantumFing}
A \textit{one-qubit fingerprinting scheme} $\mQ$ operates as follows.
To begin with, Alice and Bob receive inputs $\alpha, \beta \in S$, respectively.
Then each of them sends one qubit to the referee, who produces an output bit
based on these two qubits. There is no \textit{a priori} shared information among
the parties. Such a scheme is characterized by three components:
\begin{itemize}
\item
Alice's message set $\brace{\rho_\s}_{\sigma \in S}\,$, where $\rho_{\a}$ is the 
one-qubit (possibly mixed) state that Alice will prepare and send to the referee,
depending on her string $\alpha$.
\item
Bob's message set $\brace{\tau_\s}_{\sigma \in S}\,$, where $\tau_{\b}$ is the 
one-qubit (possibly mixed) state that Bob will prepare and send to the referee,
depending on his string $\beta$.
\item
The referee's decision strategy, which is some two-outcome measurement 
of the two-qubit state $\rho_{\a} \otimes \tau_{\b}$\,, possibly using some
finite number of ancilliary qubits.
\end{itemize}
\end{definition}

The inputs recieved by the referee fall into two classes: \emph{matching} states,
which are states of the form $\rho_\s \ox \tau_\s\,$ for some $\sigma \in S$, and
\emph{hybrid} states, which are states of the form $\rho_\a \ox \tau_\b$ for $\alpha
\ne \beta$. The goal of the referee is essentially to attempt to distinguish matching
states from hybrid states.

\subsection{Some one-qubit fingerprinting schemes}
\label{sxn:quantumEgs}

In this section, we give examples of one-qubit fingerprinting 
schemes that outperform any possible classical one-bit scheme.

The basic framework that we use for constructing fingerprinting schemes 
is that of \cite{BCWW01}: construct a set of pure
fingerprint states with small pairwise inner product and then apply a
controlled-\textsc{swap} test.

\begin{theorem}\label{bcww}
Let $\mQ$ be a fingerprinting scheme with the following characteristics:
\begin{itemize}
\item  Alice and Bob both use the same set of pure states $\brace{\ket{\phi_\s}}_{\s \in S}$ 
       as fingerprints, such that for all $\alpha, \beta \in S$
       with $\alpha \neq \beta$, $\abs{\bracket{\phi_\a}{\phi_\b}} \le \delta\,$;
\item  The referee's decision strategy is to compute the state 
       \begin{align}
	  (H \ox I \ox I) \cswap (H \ox I \ox I) 
	  \: \ket{0}\ket{\phi_\a}\ket{\phi_\b} \,,
       \end{align}
       measure the first qubit, and declare a negative result iff the result is {\bf 1}.
\end{itemize}
Then $\mQ$ is a one-sided error fingerprinting scheme, with error probability
at most $\frac{1+\delta^2}{2}$.
\end{theorem}

Using this framework for one-qubit fingerprinting schemes, the task is to find sets of
one-qubit fingerprint states such that their pairwise inner product is as small as possible.
Using the Bloch sphere representation for qubit states, this reduces to finding points
on the Bloch sphere which are as widely separated as possible. For instance, for
$\abs{S} = 4$, one would use the vertices of a regular tetrahedron inscribed within the Bloch
sphere, such as
\begin{align*}
	\ket{\phi_0} &= \ket{0}			\\
	\ket{\phi_1} &= \tfrac{1}{\sqrt 3}\paren{\ket{0} \,+\; \sqrt{2} \ket{1}} \\
	\ket{\phi_2} &= \tfrac{1}{\sqrt 3}\paren{\ket{0} \,+\; \e^{2\pi i/3}\sqrt{2} \ket{1}}\\
	\ket{\phi_3} &= \tfrac{1}{\sqrt 3}\paren{\ket{0} \,+\; \e^{4 \pi i/3}\sqrt{2}\ket{1}}
	\,.
\end{align*}
The inner product of any two of these states is $\frac{1}{\sqrt 3}$ in absolute value.
Similarly, for $\abs{S} = 3$ one would use the vertices of an equilateral triangle, and
for $\abs{S} = 6$ one would use the vertices of a regular octahedron. The (maximal)
absolute value of the inner product of two states are $\frac{1}{2}$ and $\frac{1}{\sqrt 2}$\,,
respectively, in the latter two cases. For $\abs{S} = s \gg 1$, there exists
a collection of $s$ states whose maximal pairwise inner product is $1 - O(\frac{1}{s})$
in absolute value (this may be shown using a simple geometric argument, by considering
the area available to be used in distributing the states on the Bloch sphere).

Using these figures for the magnitude of the inner product in various cases, we may use
Theorem~\ref{bcww} to construct fingerprinting schemes with one sided error, each of
which outperforms any possible one-bit fingerprinting scheme with one-sided error. The
error probabilities for these schemes are presented in Table~\ref{table:qmErrorProbabilities}, 
below.

Any one-sided error fingerprinting scheme with error probability $\epsilon$ can be
converted into a general (two-sided error) scheme with error probability
$\frac{\varepsilon}{1+\epsilon}$\,,  as follows. \label{discn:balancing} After
running the one-sided error scheme, if the result is positive, we change the
result to negative with probability $\frac{\varepsilon}{1+\varepsilon}$\,. It is
straightforward to verify that the resulting (two-sided) error probability will be
$\frac{\varepsilon}{1+\varepsilon}$\,. Using this observation, we can determine the
error probabilties for one-qubit fingerprinting schemes with two-sided error,
each of which again outperforms any possible one-bit fingerprinting scheme with
two-sided error. These error probabilities are also presented in
Table~\ref{table:qmErrorProbabilities}.

\begin{table}[htbp]
  \begin{center}
    \renewcommand{\arraystretch}{1.5}
    \begin{tabular}{|c|c|c|c|}
      \hline 
                & Maximal        & Error probability & Error probability \\[-3.7mm]
      $\abs{S}$ &                &                   &                   \\[-3.7mm]
                & inner product  & (1-sided error)   & (2-sided error)   \\
      \hline
      $3$ & $\tfrac{1}{2}$             
          & $\tfrac{5}{8} = 0.625$       & $\tfrac{5}{13} \approx 0.385$ \\
      $4$ & $\tfrac{1}{\sqrt 3}$       
          & $\tfrac{2}{3} \approx 0.667$ & $\tfrac{2}{5} = 0.400$  \\
      $6$ & $\tfrac{1}{\sqrt 2}$       
          & $\tfrac{3}{4} = 0.750$       & $\tfrac{3}{7} \approx 0.429$  \\
      $s \gg 1$ & $1 - O(\tfrac{1}{s})$ &
            $1 - O(\tfrac{1}{s})$ & $\tfrac{1}{2} - O(\tfrac{1}{s})$ \\
      \hline
    \end{tabular}
    \caption{
    \label{table:qmErrorProbabilities}
    For $\abs{S} = 3$, $4$, $6$, and the asymptotic case $s \gg 1$, the error
    probabilities for some one-qubit fingerprinting schemes in the framework
    of \cite{BCWW01}.
    }
    \end{center}
\end{table}

Note that the one-sided error probabilities are all strictly less than $1$, and the
two-sided error probabilities are all strictly less than $\tfrac{1}{2}$, neither of
which are possible for one-bit fingerprinting schemes (by Corollary~\ref{thm:classicalOneSide}
and Corollary~\ref{thm:classicalTwoSide} respectively).

\subsection{Error probabilities in one-qubit fingerprinting schemes}

In this section, we give some upper and lower bounds on the error probabilities of
one-qubit fingerprinting schemes with one-sided error. In doing so, we obtain results
which imply that when Alice's choice of fingerprint states is evenly distributed on the Bloch
sphere, the optimal one-qubit scheme is symmetric (that is, where Alice and Bob use
the same fingerprint states). We also show that there are choices of fingerprint states
for Alice where a symmetric scheme is \emph{not} optimal, and discuss necessary conditions
which must hold for a symmetric scheme to be a sub-optimal choice. These results
suggest that for even such a small instance of quantum fingerprinting, the question of
finding an optimal scheme for a fixed value $\abs{S}$ is not simple: in particular,
analysis in the general case does not lend itself to simple arguments by symmetry.

It will be useful to define some additional notation. Throughout this section, for
a one-qubit fingerprinting scheme $\mQ$, we represent the worst-case probability of a
false negative by $W_\mQ^-$, and the worst-case probability of a false positive by $W_\mQ^+$.

\subsubsection{One-qubit fingerprinting with one-sided error}
\label{sxn:strictSchemes}

As we saw in section~\ref{sxn:classical}, a one-\emph{bit} fingerprinting scheme with
one-sided error will have an error probability of~1. Because we have no interest in
one-qubit schemes with this error probability, we will use the following terminology:

\begin{definition}
A fingerprinting scheme is \emph{strict} if it has one-sided error
($W_\mQ^- = 0$) and an error probability strictly less than 1 ($W_\mQ^+ < 1$).
\end{definition}

In a strict one-qubit fingerprinting scheme $\mQ$, the referee's decision strategy
\emph{must} accept any matching state $\rho_\s \ox \tau_\s$ (for $\sigma \in S$).
This leads to a natural definition:

\begin{definition}
The \emph{accept space} $A_\mQ$ of a one-qubit fingerprinting scheme $\mQ$ is the set of
two-qubit pure states $\ket{\gamma}$, such that the referee's decision strategy would
produce a positive result with certainty if it were provided the input $\ket{\gamma}$
(regardless of whether $\ket{\gamma}$ is a product of fingerprint states from Alice and
Bob).
\end{definition}

The accept space is a vector space: if the referee's strategy accepts two distinct
two-qubit states $\ket{\gamma}$ and $\ket{\gamma'}$ \emph{with certainty}, by linearity
it must accept any linear combination of $\ket{\gamma}$ and $\ket{\gamma'}$.

In order to be accepted with certainty, each matching state must be a (possibly trivial)
mixture of states in the accept space. So, it is natural to try to understand a
fingerprinting scheme $\mQ$ in terms of $A_\mQ$\,. One question which bears interest
is how large the accept space can be for a strict one-qubit fingerprinting scheme.

\begin{theorem}
\label{thm:dim3}
If $\mQ$ is a strict one-qubit fingerprinting scheme, then $\dim A_\mQ = 3$.
\end{theorem}
\begin{proof}[sketch]
First, it is useful to note that neither Alice nor Bob can use one fingerprint state
for two distinct data strings: this would lead immediately to the existence of a
matching state (which \emph{must} be accepted with certainty) which is identical
with some hybrid state (which must \emph{not} be accepted with certainty). So, each
of Alice's fingerprint states must be distinct from one another, and similarly for Bob.
We may then prove the theorem by eliminating every other possible value of $\dim A_\mQ$.

The fact that both Alice and Bob must use distinct fingerprints for distinct input strings
implies $\dim A_\mQ > 1$\,. Using arguments of linear independence, it is possible to show
that if Alice or Bob use mixed states for fingerprints, then $\dim A_\mQ > 2$\,. On the other
hand, if Alice uses the pure states $\ket{\phi_\s}$ and Bob uses the pure states
$\ket{\psi_\s}$ for $\sigma \in S$, we can determine that $\dim A_\mQ > 2$ from the
fact that the equation
\begin{align*}
     \ket{\phi_\omega}\ket{\psi_\omega} \;=\; c_1 \ket{\phi_\mu}\ket{\psi_\mu}
                                        + c_2 \ket{\phi_\nu} \ket{\psi_\nu}
\end{align*}
cannot be satisfied for any $c_1\,, c_2 \in \C$ for any three strings $\mu, \nu, \omega \in S$
(the state on the left-hand side being both a product state and linearly independent from both
of the states on the right-hand side individually). Finally, because there must be a non-zero
probability of rejecting hybrid states, we must have $\dim A_\mQ < 4$\,. Thus we have
$\dim A_\mQ = 3$, as required.
\end{proof}

This restriction on the nature of $A_\mQ$ has some strong implications about what
states Alice and Bob may use for fingerprints. The simplest is as follows:
\begin{theorem}
\label{thm:strictNoMixed}
In a strict one-qubit fingerprinting scheme, Alice and Bob \emph{must} use pure states for
fingerprint states.
\end{theorem}
This result is easy to demonstrate: the proof can be found in the appendix.
On the other hand, this result is not clearly trivial --- without this theorem, there
is no immediately obvious reason to exclude mixed quantum fingerprinting schemes, even
for reasons of optimality. (In the classical scheme where we allow arbitrary length
fingerprints, \cite{Amb96} provides a mixed fingerprinting strategy which provides
optimal performance for a classical fingerprinting scheme.) In consideration of
Theorem~\ref{thm:strictNoMixed}, we will use the pure state
$\ket{\phi_\s}$ to describe Alice's fingerprint for $\sigma \in S$, and
$\ket{\psi_\s}$ to describe Bob's fingerprints, for $\sigma \in S$, in any
strict one-qubit fingerprinting scheme.

For a strict scheme $\mQ$, we may consider the set of two-qubit
states which are perfectly distinguishable from the matching states. Because the accept
space defined by the matching states is of dimension 3, there will
be precisely one such state, which we denote by $\ket{R_\mQ}$. Being orthogonal
to the accept space, it is the two-qubit state which the referee's decision strategy
would be most likely to reject. As a result, we call $\ket{R_\mQ}$ the \emph{reject state}
of $\mQ$.\label{discn:rejectState} Although it is possible for the referee's strategy to
accept the state $\ket{R_\mQ}$ with non-zero probability, an optimal choice of decision
strategy for the referee is to reject $\ket{R_\mQ}$ with certainty.

\begin{theorem}
\label{thm:optimalStrictDecision}
Let $C$ be a (non-empty) collection of strict one-qubit fingerprinting schemes,
characterized by some choice of fingerprint states $\brace{\ket{\phi_\s}}_{\s \in S}$ for
Alice and $\brace{\ket{\psi_\s}}_{\s \in S}$ for Bob. Let $A$ be the accept space
defined by these fingerprint states, and $\ket{R}$ the corresponding reject state.
Then for any $\mQ \in C\,$,
\begin{align}
    W_\mQ^+ \;\;\ge\;\;
    1 \;-\; \min_{\alpha \ne \beta}\,
    \abs{\bra{R} \paren{\Big. \ket{\phi_\a} \ox \ket{\psi_\b} \Big.}}^2 \,,
\end{align}
with equality when the referee's decision strategy is to project $\ket{\phi_\a}\ket{\psi_\b}$
either into $A$ or onto $\ket{R}$ using a measurement, returning a negative
result iff the result of this measurement is $\ket{R}$.
\end{theorem}
\begin{proof}
That the strategy described above yields a strict one-qubit fingerprinting scheme
is trivial: we will show that this is the optimal decision strategy for the referee.
Whatever the referee's decision strategy may be, it must attempt to distinguish between
the hybrid states and the accept space without having specific knowledge about the input
he receives. Any such strategy can be described by a two-valued POVM $\brace{E_+\,,\,E_-}\,$,
where $E_+$ and $E_-$ corresponds positive and negative results respectively.
For any state $\ket{\psi} \in A$, we require 
\begin{align}
    \tr\ (E_- \densmtx{\psi}) &\,=\, 0 \,,
\end{align}
which implies  $E_- = p \densmtx{R}$ for some probability $p \ge 0$. Then, the POVM
which describes the referee's strategy is equivalent to a partial projective
measurement (either onto the the state $\ket{R}$ or into the space $A$), followed
by a biased coin-flip to decide whether to produce a negative result should the state
collapse to $\ket{R}$. In order to achieve the greatest probability of distinguishing
hybrid states from the accept space, we may choose to fix $p = 1$. The strategy that
we obtain is then precisely the one described by the theorem.
\end{proof}

\label{discn:Schmidt}
The reject state can be a useful tool for analysis. For instance, consider the Schmidt
decomposition of $\ket{R_\mQ}$, which can be expressed as
\begin{align}
    \ket{R_\mQ} \;&=\; \tfrac{1}{\sqrt{1 + C^2}}
                       \paren{\ket{\eta_0} \ket{\kappa_0} \,-\, C \ket{\eta_1} \ket{\kappa_1}}
\end{align}
for some orthonormal bases $\brace{\ket{\eta_0}\,, \ket{\eta_1}}$ and
$\brace{\ket{\kappa_0}\,, \ket{\kappa_1}}$ of $\C^2$, and some real coefficient
$C > 0$\,. (Because Alice and Bob are not allowed to use the same fingerprint
state for different strings, it is simple to show that $C \ne 0$\,, and we may absorb
any relative phase into the definitions of $\ket{\eta_1}$ and $\ket{\kappa_1}$\,.) Let
$U = \ketbra{0}{\eta_0} + \ketbra{1}{\eta_1}$ and $V = \ketbra{1}{\kappa_0}
+ \ketbra{0}{\kappa_1}$\,. Then, we have
\begin{align}
    \label{eqn:canonicalRejectState}
    (U \ox V) \ket{R_\mQ} \;&=\; \tfrac{1}{\sqrt{1 + C^2}}
                       \paren{\ket{01} \,-\, C \ket{10}}
\end{align}
As a part of his decision strategy, the referee can perform the transformation $U \ox V$
as the first step, without loss of generality. However, because $U \ox V$ can also be
carried out without communication by Alice and Bob as the last part of the
preparation of their fingerprint states, $\mQ$ is equivalent to a different one-qubit
scheme $\mQ'$, where Alice and Bob use the fingerprints $U \ket{\phi_\s}$ and
$V \ket{\psi_\s}$ (respectively) for the string $\sigma \in S$\, rather than $\ket{\phi_\s}$
and $\ket{\psi_\s}$\,, and where the reject state is $(U \ox V) \ket{R_\mQ}$ rather
than $\ket{R_\mQ}$. For the sake of brevity, we will call a strict scheme $\mQ$
\emph{canonical} if the reject state is given by $\ket{R_\mQ} = \tfrac{1}{\sqrt{1 + C^2}}
\paren{\ket{01} - C \ket{10}}$ for some $C  > 0$. For any strict scheme $\mQ$\,, we will
call the canonical fingerprinting scheme which is equivalent to $\mQ$ the \emph{canonical
form} of $\mQ$.

Given the canonical form of $\mQ$\,, we can easily describe a relation between Alice's
and Bob's fingerprint states. We can express $\ket{\phi_\s}$ and $\ket{\psi_\s}$ in the
form
\begin{align}
  \label{eqn:complexExpression}
  \begin{split}
     \ket{\phi_\s} \;&=\;  \tfrac{1}{\sqrt{1 + \abs{u_\s}^2}}
                         \paren{\ket{0} \Big.+ u_\s\ket{1}}     \\
     \ket{\psi_\s} \;&=\; \tfrac{1}{\sqrt{1 + \abs{w_\s}^2}}
			 \paren{\ket{0} \Big.+ w_\s\ket{1}}\,,
  \end{split}
\end{align}
for some complex constants $u_s\,,w_s \in \C$. (For $\ket{\phi_\s} = \ket{1}$\,, we may take
the limit $\abs{u_\s} \smallto \infty$\,, and similarly for $\ket{\psi_\s}$\,.)\footnote{To
ensure that this is meaningful, we take care to observe that expressions involving
$\abs{u_\s}$ and $\abs{w_\s}$ extend in an appropriate way under this limit.}
Using the expression in Equation~\ref{eqn:canonicalRejectState} for $\ket{R_\mQ}$\,,
from $\bra{R_\mQ}\paren{\ket{\phi_\s} \ox \ket{\psi_\s}} = 0$, we can then determine
\begin{align}
   \label{eqn:uwLinRelation}
   \begin{split}
       w_\s \bracket{01}{01}& - C u_\s \bracket{10}{10} \;=\; 0
       \\ \implies& \quad w_\s \;=\; C u_\s\,.
   \end{split}
\end{align}
(This equation extends appropriately in the limit $\abs{u_\s}, \abs{w_\s} \smallto \infty$
in the sense that $\ket{\phi_\s}\ket{\psi_\s}$ will approach $\ket{11}$\,, which is an element
of $A_\mQ$.) Then, if the reject state $\ket{R_\mQ}$ of a strict one-qubit scheme $\mQ$ is
known, one can determine Bob's fingerprints from Alice's fingerprints, using the canonical
form of $\mQ$. One can then characterize a strict one-qubit fingerprinting protocol
by its' reject state $\ket{R_\mQ}$\,, the probability with which the referee rejects
$\ket{R_\mQ}$\,, and Alice's choice of fingerprint states.

Using this observation, we can also determine useful information about a strict one-qubit
scheme from only Alice's or only Bob's fingerprint states, along with the parameter $C$
used to define the reject state.
\begin{theorem}
\label{thm:Kconst1}
Let $\mQ$ be a strict one-qubit fingerprinting scheme, where Alice's fingerprint states
are given by $\brace{\ket{\phi_\s}}_{\s \in S}$ and Bob's fingerprint states are given by
$\brace{\ket{\psi_\s}}_{\s \in S}\,$. For the sake of brevity, define the hybrid
state $\ket{h_{\a\b}} = \ket{\phi_\a}\ket{\psi_\b}$ for each $\alpha \ne \beta$.
Then, there is a collection of positive real constants $\brace{K_\s}_{\s \in S}$
such that
\begin{align}
          \label{eqn:Kconst}
          \abs{\bracket{R_\mQ}{h_{\a\b}}}^2
	  \;=\; K_\b \paren{1 \;-\; \abs{\bracket{\phi_\b}{\phi_\a}}^2} \,,
\end{align}
for all $\alpha, \beta \in S$.
\end{theorem}
With this result, knowledge of the constants $K_\s$ for $\sigma \in S$ allows
one to easily determine the probability that any hybrid state will be projected
onto $\ket{R_\mQ}$\,, using only Alice's fingerprints. The proof of
Theorem~\ref{thm:Kconst1} is left to the appendix; for schemes $\mQ$
in canonical form, the parameter $K_\sigma$ can be given as a function
of $C$, and the extended real number $u_\sigma$ which defines Alice's state
$\ket{\phi_\s}$\,: 
\begin{align}
       \label{eqn:canonicalK}
       K_\s = \frac{C^2 \big(1 + \abs{u_\s}^2\big)}
                   {\big(1 + C^2 \abs{u_\s}^2\big)\big(1 + C^2\big)}
\end{align}
(As $\abs{u_\s} \smallto \infty$\,, this formula converges to $\frac{1}{1 + C^2}$\,.)
It is easy to show that, as a function of $\abs{u_\s}$\,, $K_\s$ is constant for
$C = 1$\,, monotone increasing for $C < 1$\,, and monotone decreasing for $C > 1$\,.
Of particular interest is the locus of possible values of $u_\s$ where
$K_\s = \frac{1}{2}$\,: it is simple to show that for any value of $C$, this
set of points will be the circle of radius $C\inv$ about the origin.

By considering this formula for $K_\s$\,, we can begin to try finding an
optimal strict fingerprinting scheme (by optimising the value of $C$)
based on Alice's fingerprint states, and vice-versa. We will describe
some ideas of how this may be done in the following section.

\subsubsection{Strict symmetric fingerprinting schemes}

Fingerprinting schemes where Alice and Bob use the same states as each other
for their fingerprints are a natural class of fingerprinting schemes to consider.
All of the results obtained in the previous section hold for the special
case of strict symmetric one-qubit fingerprinting schemes: in this section, we will
elaborate on some of the results for this special case, and discuss when
a symmetric scheme is an optimal choice of one-qubit fingerprinting scheme.

It is easy to verify that, for a symmetric scheme, $\ket{R_\mQ} = \ket{\Psi^-}$.
Theorem~\ref{thm:optimalStrictDecision} then effectively reproduces the
controlled-\textsc{swap} decision strategy of Theorem~\ref{bcww}, as $\ket{\Psi^-}$
is a $-1$ eigenstate of the \textsc{swap} operation, and the other eigenstates of
\textsc{swap} have the eigenvalue $+1$. As well, $\ket{R_\mQ} = \ket{\Psi^-}$ gives
us $C = 1$ for $\mQ$\,, which implies that $K_\s = \frac{1}{2}$ for all $\sigma \in S$.
Theorem~\ref{thm:Kconst1} then simplifies to
\begin{align*}
      \abs{\bracket{R_\mQ}{h_{\a\b}}}^2 = \frac{1 - \abs{\bracket{\phi_\b}{\phi_\a}}^2}{2}\,.
\end{align*}
The probability of rejection obtained is then consistent with the results of \cite{BCWW01}.

Considering $K_\s$ as a function of $C$ and $u_\s$\,, the value $C = 1$ is the only
value for which $K_\s$ is not strictly increasing or strictly decreasing. As a result,
the choice $C = 1$ plays a somewhat special role in the analysis of strict one-qubit
schemes. One approach to examining one-qubit fingerprinting schemes is to fix Alice's
distribution of fingerprints, and try to determine which value of $C$ leads to the best
error probability for the scheme as a whole. Under some reasonable assumptions on Alice's
choice of states, the value $C = 1$ (and thus, a symmetric scheme) is optimal. However,
this is not true under more general assumptions.

Suppose Alice has some fixed choice of fingerprinting states. Among Alice's fingerprints,
there will be some pair of states $\ket{\phi_\a}\,,\,\ket{\phi_\b}$ which have the largest
inner product of any pair. For one-qubit schemes in canonical form, in order for an asymmetric
scheme to perform better than a symmetric scheme, we require $K_\a\,, K_\b > \frac{1}{2}$ for
this pair of $\alpha$ and $\beta$. This is possible only if both $u_\a$ and $u_\b$ lie within
a circle of radius $C\inv$ about the origin on the complex plane for $C > 1$, or outside such
a circle for $C < 1$. Either of these constraints on the values of $u_\a$ and $u_\b$ is
equivalent to the states $\ket{\phi_\a}$ and $\ket{\phi_\b}$ lying within some circle of
lattitude, either closer to $\ket{0}$ or to $\ket{1}$, on the Bloch sphere. Then, the closer
such a minimally distinguishable pair lies to the equator of the Bloch sphere, the closer
$C$ must be to 1 for these criteria to apply, and so the closer the optimal scheme will be
to symmetric.

Using this observation, we can determine some necessary criteria on Alice's choice
of fingerprints, in order for a symmetric scheme to be \emph{suboptimal} for that choice.
For a minimally distinguishable pair of states $\ket{\phi_\a}$ and $\ket{\phi_\b}$ of Alice's,
if
\begin{itemize}
\item one of the pair lies on the equator of the Bloch sphere,
\item the two states lie on opposite sides of the Bloch equator, or
\item two such pairs of minimal distinguishability lie on opposite sides of the Bloch equator,
\end{itemize}
then it is easy to show that a symmetric protocol is optimal. (In particular, for a
scheme where Alice's states are optimally spaced on the Bloch sphere, the latter two
conditions will most likely apply.) In order for $C > 1$ to be optimal, we then require
$u_\a$ and $u_\b$ to both lie within a circle of radius $C\inv$ on the complex plane, and
for $\abs{\bracket{\phi_\a}{\phi_\b}}$ to be \emph{strictly} larger than the inner product
of any other pair. For $C < 1$ to be optimal, we have similar requirements, except that
$u_\a$ and $u_\b$ must both lie outside the circle of radius $C\inv$ on the complex plane.

It is easy to find simple choices of fingerprint states
for Alice where a symmetric scheme is not optimal. One example would be where
$S = \brace{0,1,2}$, and we define Alice's states by the complex parameters
\begin{align*}
       u_{\sss 0} = 0\,;  \quad
       u_{\sss 1} = 2\,;  \quad
       u_{\sss 2} = -2\,.
\end{align*}
In this case, the minimally distinguishable pair would be $\ket{\phi_1}$ and $\ket{\phi_2}$
with an inner product of $\frac{3}{5}$, compared to $\frac{1}{5}$ for the other two possible
pairs. Using Theorem~\ref{thm:Kconst1} with the value $K_\s = \frac{1}{2}$ for all
$\sigma \in S$,  a symmetric scheme $\mQ$ with this choice of fingerprints for Alice would
have error probability $\frac{17}{25} = 0.68$\,. However, the points $u_{\sss 1}$ and
$u_{\sss 2}$ both lie outside the circle of radius $C\inv$ on the complex plane, for
any $\frac{1}{2} < C < 1$. In particular, consider a scheme $\mQ_A$ where
$C = \frac{1}{\sqrt 2}$. This choice of $C$ will yield
\begin{align*}
       K_0 = \tfrac{1}{3}\,; \quad
       K_1 = K_2 = \tfrac{5}{9}
\end{align*}
The higher values for $K_1$ and $K_2$ improve the probability of error in the worst case
at the cost of the probability of success in other cases. Using Theorem~\ref{thm:Kconst1}
once more, we may determine the probability of error when $\alpha = 1$ and $\beta = 2$
to be $\frac{29}{45} \approx 0.644$\,. It is easy to determine that this is the
error probability of $\mQ_A$. Although this is not an improvement on the error probability
of the strict protocol for $\abs{S} = 3$ presented in Table~\ref{table:qmErrorProbabilities},
this is better than in the symmetric protocol $\mQ_S$. Then, for this choice of
fingerprint states for Alice, the scheme $\mQ_A$ has a better worst-case probability
of error than a symmetric scheme.

In general, evenly spacing Alice's states on the Bloch sphere will lead to a symmetric
fingerprinting scheme being the optimal choice, because there will not be a unique
pair with maximal inner product, and the distribution of minimally distinguishable pairs
will be essentially symmetric on the Bloch sphere. However, for an arbitrary distribution
of Alice's states, a symmetric protocol is not optimal, and it appears difficult to show that
evenly spacing states on the Bloch sphere is an optimal choice without first assuming a
symmetric scheme is to be used.

\section{Conclusion}

We have shown that fingerprinting schemes that make use of only one-qubit messages
from Alice and Bob to the referee can perform better in the worst case than is
possible for any similar classical communication scheme using single bit
messages. This result holds whether one requires one-sided error (with no false
negatives) or one allows two-sided error probabilities.

For a one-bit fingerprinting scheme, one can never achieve a better worst-case
performance than blind guessing, and one-sided error schemes have error probability
$1$. The number of inputs for Alice and Bob where the fingerprinting scheme performs
``badly'' (with a high lower bound on the error probability) either includes all inputs
where Alice's and Bob's string match, or a sizeable fraction of the other possible inputs.

In the quantum case, while some lower bounds exist, upper bounds for error probability
can also be easily derived for one-qubit fingerprinting schemes with one-sided error,
and can be determined from the fingerprint states of either Alice or Bob alone
if the reject state is known. Using these techniques, it is possible to show that
there are choices of fingerprints for Alice, for which the optimal one-qubit scheme
is not symmetric. However, if it is assumed that Alice distributes her states as
evenly over the Bloch sphere as possible, a symmetric scheme is quite likely the
optimal scheme.

It is more difficult to make clear statements one-qubit fingerprinting schemes with
two-sided error. Although it is easy to find ones with reasonably low error probability
by converting one-sided error schemes, the structure of strict schemes is lost: the
freedom to reject matching states with some probability makes simple arguments by
linearity difficult to find.

\section*{Acknowledgements}
I would like to thank Richard Cleve for suggesting this problem, providing
initial insights, and useful critical advice on this article.

\appendix
\section*{Appendix: Deferred proofs}
\label{apx}

We will now present the two results from the analysis of strict one-qubit fingerprinting
schemes whose proofs were deferred. 

To prove Theorem~\ref{thm:strictNoMixed}, we will use a number of ideas introduced
originally after the theorem: namely, the idea of the reject state (defined on
page~\pageref{discn:rejectState}), and the idea of the canonical form of a
strict protocol $\mQ$ (presented on page~\pageref{discn:Schmidt}). These two
ideas do not depend on Alice's and Bob's fingerprint states being pure states.
We would also like to use the linear relation of Equation~\ref{eqn:uwLinRelation},
which we will briefly re-derive here.

~\\[-4mm]
\begin{proof}[of Theorem~\ref{thm:strictNoMixed}]
Without loss of generality, $\mQ$ is in canonical form. Then, let $C$ be a positive real
such that
\begin{align*}
    \ket{R_\mQ} \;&=\;  \tfrac{1}{\sqrt{1 + C^2}}
                        \paren{\ket{01} - C\ket{10}}
\end{align*}
holds. Next, consider a product state $\ket{u}\ket{w} \in A_\mQ$\,.
We may represent the state $\ket{u}$ and $\ket{w}$ in the form
\begin{align*}
   \ket{u} \;&=\; \tfrac{1}{\sqrt{1 + \abs{u}^2}} \paren{\ket{0} \Big.+ u\ket{1}}\\
   \ket{w} \;&=\; \tfrac{1}{\sqrt{1 + \abs{w}^2}} \paren{\ket{0} \Big.+ w\ket{1}}\,,
\end{align*}
for some complex constants $u,w \in \C$. As $\bra{R_\mQ}\paren{\ket{u} \ox \ket{w}} = 0$,
we can then determine
\begin{align*}
   w \bracket{01}{01} - C u \bracket{10}{10} \;=\; 0
   \\[-8mm]
\end{align*}
\begin{align*}
   \implies \quad w \;=\; C u\,.
\end{align*}
Considering this, let $\rho_\s \ox \tau_\s$ be some matching state of $\mQ$.
Suppose that Alice uses a mixed state for $\rho_\s\,$. Then, let $\rho_\s$ be
some mixture of the states
\begin{align*}
      \ket{\phi_\s} \;&=\; \tfrac{1}{\sqrt{1 + \abs{u_\s}^2}}
	           \paren{\ket{0} + u_\s \ket{1}}               \\
      \ket{\phi'_\s} \;&=\;\tfrac{1}{\sqrt{1 + \abs{u'_\s}^2}}
	           \paren{\ket{0} + u'_\s \ket{1}}\,,
\end{align*}
and let $\tau_\s$ be a (possibly trivial) mixture including the state
\begin{align*}
      \ket{\psi_\s} \;&=\; \tfrac{1}{\sqrt{1 + \abs{w_\s}^2}}
	           \paren{\ket{0} + w_\s \ket{1}}
\end{align*}
Because $\ket{\phi_\s}\ket{\psi_\s}$ and $\ket{\phi'_\s}\ket{\psi_\s}$ are both
in $A_\mQ$\,, it must be that $w_\s = C u_\s$ and $w_\s = C u'_\s$\,. This is only
possible if $u_\s = u'_\s$\,, so that $\ket{\phi_\s} = \ket{\phi'_\s}$. Then,
$\rho_\s$ is a pure state. A similar analysis holds for Bob.
\end{proof}

Next, we present the proof of Theorem~\ref{thm:Kconst1}. Implicit in the proof
for the case where $u_\a\,,u_\b \in \C$\, is a derivation of the formula
for $K_\s$ for canonical schemes that was shown previously, in Equation~\ref{eqn:canonicalK}.

~\\[-4mm]
\begin{proof}[of Theorem~\ref{thm:Kconst1}]
Without loss of generality, let $\mQ$ be in canonical form. Let $\alpha \ne \beta$, and let
\begin{align*}
    \ket{R_\mQ} \;&=\; \tfrac{1}{\sqrt{1 + C^2}}
                    \paren{\ket{01} - C\ket{10}} \,,
\end{align*}
as before. We may express $\ket{\phi_\a}$ and $\ket{\phi_\b}$ in terms of (extended)
complex numbers, as in Equation~\ref{eqn:complexExpression}: for instance, we have
\begin{align*}
    \ket{\phi_\a} \;&=\; \tfrac{1}{\sqrt{1+\abs{u_\a}^2}}
                         \paren{\ket{0} + u_\a \ket{1}} \,,
\end{align*}
and similarly for $\ket{\phi_\b}$. In the case where $\ket{\phi_\a} = \ket{1}$, we set
$u_\a = \infty$ (using the one-point compactification of $\C$). We may then
use Equation~\ref{eqn:uwLinRelation} to determine Bob's fingerprint states: for
instance,
\begin{align*}
    \ket{\psi_\a} \;&=\tfrac{1}{\sqrt{1 + C^2 \abs{u_\a}^2\;}}
                \paren{\ket{0} \Big.\:+\: C u_\a \ket{1}} \,,
\end{align*}
and similarly for $\ket{\psi_\b}$. Then, we will show that Equation~\ref{eqn:Kconst}
is satisfied for
\begin{align}
      \label{eqn:KsigmaFormula}
      K_\s \;=\; \frac{C^2 \big(1 + \abs{u_\s}^2\big)}
                      {\big(1 + C^2 \abs{u_\s}^2\big)\big(1 + C^2\big)} \;.
\end{align}
In the case $u_\s = \infty$, this expression has a well-defined
limit of $K_\s = \frac{1}{1 + C^2}$.

We have three cases: $u_\a = \infty$ (in which case $u_\b \in \C$), $u_\b = \infty$
(in which case $u_\a \in \C$), and $u_\a\,,u_\b \in \C$. The first two cases are
easy to prove; the last case of $u_\a\,,u_b \in \C$ requires a little more analysis.
Consider the value of
$\abs{\bracket{\phi_\b}{\phi_\a}}^2\,$:
\begin{align*}
          \abs{\bracket{\phi_\b}{\phi_\a}}^2 \;&=\;
	  \tfrac{\abs{1 + u_\b\conj u_\a}^2}
	       {\paren{1 + \abs{u_\a}^2}\paren{1 + \abs{u_\b}^2}} \,.
\end{align*}
By subtracting both sides of the equation from $1$, we obtain
\begin{align*}
          1 \;-\; &\abs{\bracket{\phi_\b}{\phi_\a}}^2    \\
	  &=\;\;
	  \tfrac{\paren{1 + u_\a\conj u_\a}\paren{1 + u_\b\conj u_\b} \;\;-\;\;
	        \paren{1 + u_\b\conj u_\a}\paren{1 + u_\a\conj u_\b}}
		{\paren{1 + \abs{u_\a}^2}\paren{1 + \abs{u_\b}^2}}    \\
	  &=\;\;
	  \tfrac{\abs{u_\a - u_\b}^2}{\paren{1 + \abs{u_\a}^2}\paren{1 + \abs{u_\b}^2}}\,,
\end{align*}
from which we conclude
\begin{align}
          \label{eqn:diff_ua_ub}
	  \begin{split}
	      &\tfrac{\abs{u_\a - u_\b}^2}{1 + \abs{u_\a}^2} \;\;=\;\;
	      \paren{1 + \abs{u_\b}^2}\paren{1 \;-\; \abs{\bracket{\phi_\b}{\phi_\a}}^2} \,.
	  \end{split}
\end{align}
We will make use of this equality in our analysis of $\abs{\bracket{R_\mQ}{h_{\a\b}}}$.
We may expand the expression for $\ket{h_{\a\b}}$ in terms of the standard basis:
\begin{align*}
          \ket{h_{\a\b}} \;=\;
	  \tfrac{\ket{00} \;+\; C u_\b \ket{01}
	         \;+\; u_\a \ket{10} \;+\; C u_\a u_\b \ket{11}}
		{\sqrt{1 + \abs{u_\a}^2\;}\sqrt{1 \;+\; C^2 \abs{u_\b}^2\;}} \;\;.
\end{align*}
Taking the inner product with $\ket{R}$, and squaring the absolute value,
we obtain
\begin{align}
        \label{eqn:part_way_upper_bound}
          \abs{\bracket{R_\mQ}{h_{\a\b}}}^2 \;&=\;
	  \tfrac{C^2 \abs{u_\a - u_\b}^2}
	  {\paren{1+\abs{u_\a}^2}\paren{1 \;+\; C^2 \abs{u_\b}^2\;}
	  \paren{1 + C^2}} \;\;.
\end{align}
Applying Equation~\ref{eqn:diff_ua_ub}, we can perform a substitution to obtain
\begin{align*}
          \abs{\bracket{R}{h_{\a\b}}}^2 \;&=\;
	  \tfrac{C^2 \paren{1 + \abs{u_\b}^2}}
	  {\paren{1 \;+\; C^2 \abs{u_\b}^2\;}
	  \paren{1 + C^2}}\;
	  \paren{1 \;-\; \abs{\bracket{\phi_\b}{\phi_\a}}^2}
	  \\[2mm] &=\;
	  K_\b  \paren{1 \;-\; \abs{\bracket{\phi_\b}{\phi_\a}}^2} \,.
\end{align*}
The theorem then holds.
\end{proof}

\bibliographystyle{amsalpha}

\end{document}